\documentclass[conference]{IEEEtran}
\IEEEoverridecommandlockouts
\usepackage{cite}
\usepackage{amsmath,amssymb,amsfonts}
\usepackage{algorithmic}
\usepackage{graphicx}
\usepackage{textcomp}
\usepackage{xcolor}
\begin{document}

\title{Optimal Decoding of Convolutional Codes using a Linear State Space Control Formulation\\

}

\author{\IEEEauthorblockN{1\textsuperscript{st} Caleb M. Bowyer}
\IEEEauthorblockA{\textit{dept. Electrical and Computer Engineering } \\
\textit{University of Florida}\\
Gainesville, USA \\
c.bowyer@ufl.edu}
}

\maketitle

\begin{abstract}
The equivalence of a systematic convolutional encoder as linear state-space control system is first realized and presented through an example. Then, utilizing this structure, a new optimal state-sequence estimator is derived, in the spirit of the Viterbi algorithm. Afterwords, a novel way to perform optimal decoding is achieved, named the Bowyer Decoder, which is a fully deterministic decoder in that the full FSM is known to the decoding algorithm.
\end{abstract}

\begin{IEEEkeywords}
Convolutional Decoding, Viterbi Algorithm, LTI Control Systems
\end{IEEEkeywords}

\section{Background and Literature Overview}
The origins of the Viterbi algorithm can be found in \cite{Viterbi67}. Later work focused on the performance of convolutional codes, \cite{Viterbi71}. Originally, Viterbi did not realize the Viterbi Algorithm was an optimal Maximum Likelihood Decoder; He thought it was sub-optimal. Read the personal accounts of the Viterbi Algorithm from Forney and Viterbi's perspective: \cite{Viterbi_personal}, \cite{Forney_personal}. A textbook treatment of the Viterbi algorithm can be found in \cite{Bert17}.

In this paper I will show how a systematic convolutional encoder has an equivalent linear state-space form, which is amenable to the concepts of control theory. The reformulation is not merely surface level, but offers new insight and angles of attack into the decoding process that the older approaches lack. The possible ways in which new decoding questions can be answered will be pointed out in the next section on the reformulation of an example convolutional encoder. 

The next section covers the form of the decoder, which operates using the mechanics of linear state-space control system. In the conclusion, I will point out future directions for this new research area into joint control and communication systems. 


\section{Reformulation of Convolutional Encoder as LTI Control System}

\begin{figure}[h]
\centering
\includegraphics[width=11cm, height=9cm]{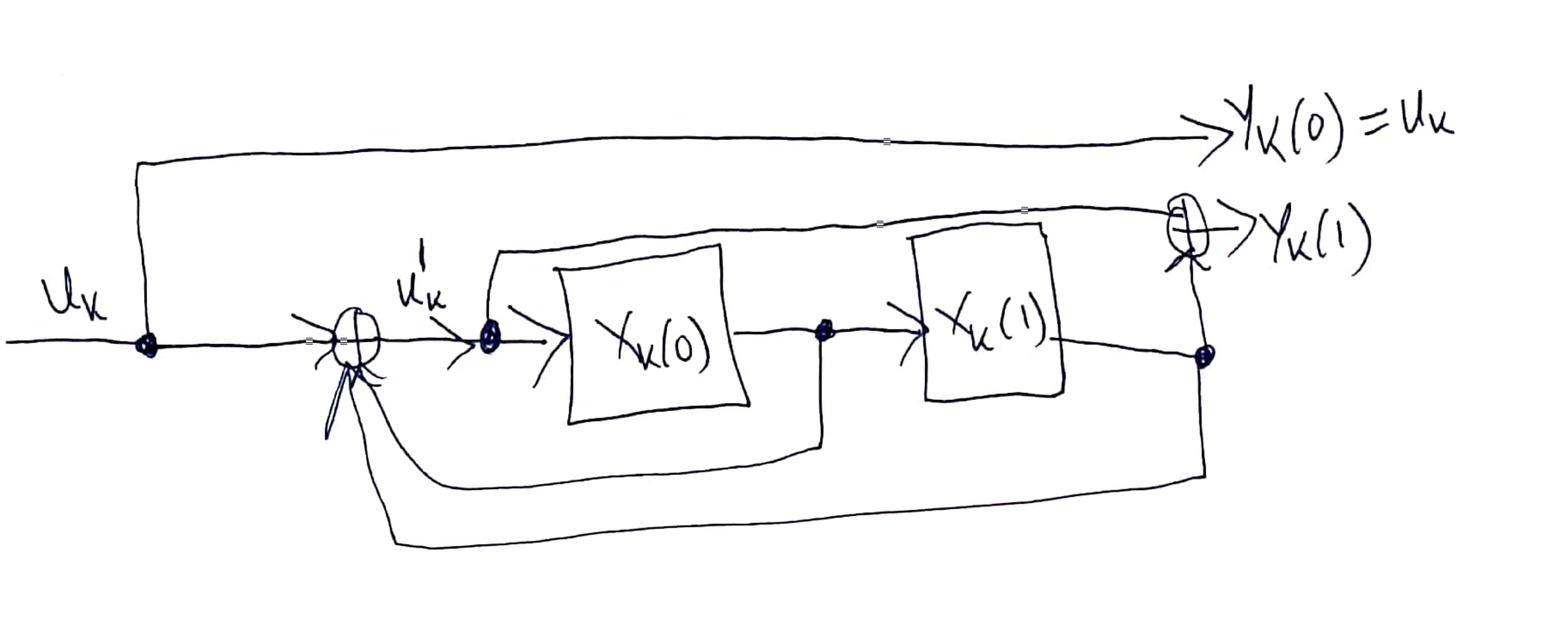}
\caption{Convolutional Encoder Reformulation Example}
\label{fig:conv}
\end{figure}

To show the equivalence between the systematic convolutional encoder and the linear state-space control system, I first present an example. The example convolutional encoder is a rate $R=\frac{1}{2}$ convolutional code where $k=1,n=2$.

See \ref{fig:conv} for how the systematic convolutional encoder is setup for this example so we can begin to break it down, and see the equivalence to an LTI control system. 

First, I should present my notation for the outputs of the encoder at some stage $k$.  Output $0$ is denoted as $y_k(0)$ on the top branch, and output $1$ is denoted as $y_k(1)$ on the bottom branch of the convolutional encoder. $\underbar{y}_k$ will be the collection of these two outputs into a single vector. 

The input bit here at some stage $k$ is $u_k$. In the diagram $u_k'$ is the IIR feedback. Each of the memory cells in convolutional encoder terminology, I view as individual state components in my LTI control system. 

Similarly to the output notation, for the first two states or "memory cells", I will use the notation $x_k(0), x_k(1)$ for the first and second state variables respectively. Then, the full state vector at some stage $k$ is $\underbar{x}_k$. 

Looking at \ref{fig:conv}, and using our understanding of LTI control systems, we can form the state evolution equations and the output equations for the encoder thus forming the LTI System parameters $A,B,C,D$. $A,B$ are for the state evolution equation, while $C,D$ are for the output equation. Any mathematical operations done with the system parameter matrices $A,B,C,D$ are carried out using mod 2 arithmetic. This is because proper codewords are formed using mod 2 arithmetic when performing encoding. Hence, the LTI control system computations are also carried our using mod 2 arithmetic in this setting. 

From the systematic encoder figure we see that



$$y_k(0)=u_k,$$ and $$ y_k(1) = x_k(1) + u_k' = x_x(1) + u_k + x_k(0) +x_k(1) = x_k(0) + u_k.$$

Putting everything into a vector output equation:

\begin{gather*}\label{obs_eqn}
 \underbar{y}_k = \begin{bmatrix}
                0 & 0\\
                1 & 0
                \end{bmatrix}
            \begin{bmatrix}
            x_k(0)\\
            x_k(1)
            \end{bmatrix}
            +
            u_k\begin{bmatrix}
            1\\
            1
            \end{bmatrix}
\end{gather*}

And the state equation can be read off from the diagram as well:

\begin{gather*}\label{state_eqn}
 \underbar{x}_k = \begin{bmatrix}
                1 & 1\\
                1 & 0
                \end{bmatrix}
            \begin{bmatrix}
            x_k(0)\\
            x_k(1)
            \end{bmatrix}
            +
            u_k\begin{bmatrix}
            1\\
            0
            \end{bmatrix}
\end{gather*}

Next, I present the table for all possible inputs $u$, recursive feedback input $u'$, current state, next state, and output for the convolutional encoder example  using the system and output matrices: 
\begin{equation*}
    A = \begin{bmatrix}
                1 & 1\\
                1 & 0
                \end{bmatrix},
\end{equation*}
\begin{equation*}
B = \begin{bmatrix}
                1 \\
                0 
                \end{bmatrix},
\end{equation*}
\begin{equation*}
C = \begin{bmatrix}
                0 & 0\\
                1 & 0
                \end{bmatrix},
\end{equation*}
\begin{equation*}
        D = \begin{bmatrix}
                1 \\
                1 
                \end{bmatrix}. 
\end{equation*}

\begin{table}[h!]
\centering
\begin{tabular}{||c c c c c||} 
 \hline
 u & u' & Current State & Next State & Output \\ [0.5ex] 
 \hline\hline
 0 & 0 & 00 & 00 & 00 \\ 
 1 & 1 & 00 & 10 & 11 \\ 
 0 & 1 & 10 & 11 & 01 \\ 
 1 & 0 & 10 & 01 & 10 \\ 
 0 & 1 & 01 & 10 & 00 \\ 
 1 & 0 & 01 & 00 & 11 \\ 
 0 & 0 & 11 & 01 & 01 \\ 
 1 & 1 & 11 & 11 & 10 \\[1ex] 
 \hline
\end{tabular}
\caption{input, state, output}
\label{table:1}
\end{table}

From this table, one can take each row as data and confirm that in all $8$ cases that the LTI control system reformulation is correct in the sense that for a specified input $u$, for a current state $x$, the next state $x'$ is produced correctly. Also, in all 8 cases the output can be confirmed as correct when performing the LTI control system computation with GF($2$) math of course, as in the state equation. 

After understanding this particular example it is easy to see how the convolutional encoder in general can be transformed into an LTI control system. The number of memory cells specifies the number of state variables. Then, one needs to determine the $A,B,C,D$ matrices based on the logical circuit. 

The goal of this reformulation is two-fold. One, looking at the error control problem from a control perspective can allow us to answer new research questions that could not have been addressed in older formulations, such as the classical Viterbi algorithm. 

As an example of this idea of combining communications and controls consider the following concept from control theory. I consider two types of illustrations, consider the state-evolution with zero-input forcing function. First, look at what happens when 
\begin{equation*}
\underbar{x}_0 = \begin{bmatrix}
                0\\
                0
\end{bmatrix}.
\end{equation*} Performing the updating to $\underbar{x}_1$ using $A$ shows the state will still be the all zeros vector. Thus, 
\begin{equation*}
\underbar{x}_k = \begin{bmatrix}
                0\\
                0
\end{bmatrix} 
\end{equation*}
$ \forall k\geq 1$. That is we know from the LTI reformulation that no matter how many times we iterate the system, we will always end up at the same state given that initial condition. 

Next, consider a different initial condition to see a different effect. After looking at this example we will be able to draw an immediate parallel between encoders for a communication system and a fundamental property of a LTI control system. 

Look at what happens when 
\begin{equation*}
\underbar{x}_0 = \begin{bmatrix}
                1\\
                1
\end{bmatrix}.
\end{equation*} Iterating the state evolution operator $A$ on the initial condition shows that we end up cycling between three states: 

\begin{equation*}
    \{\begin{bmatrix}
                1\\
                1
\end{bmatrix}, 
\begin{bmatrix}
                0\\
                1
\end{bmatrix},
\begin{bmatrix}
                1\\
                0
\end{bmatrix}\}.
\end{equation*}

Thus, this mathematical system point of view can enable quick computation of all possible cycles for any initial condition and with a zero-input forcing function. 

This reformulation also allows us to develop a notion of controllability of the convolutional encoder. Using $$\mathcal{C} = [b\vline Ab] = \begin{bmatrix}
                1 & 1\\
                0 & 1
\end{bmatrix}.$$ From this we can immediately infer the system is fully controllable because the determinant of the controllability matrix is non-zero. What this says is that any state can be reached from any other state by a certain choice of control inputs. Furthermore, control theory allows us to determine the exact control input sequence to take us from one state say $x_s$ (start state) to an ending state $x_e$ (end state). As the number of memory cells increase in the convolutional encoder, visual inspection of the trellis diagram would not be possible because it would be too densely complicated to read. Nevertheless, my method allows another more compact way to find any possible state path's corresponding input sequence using the LTI control system formulation. 

Other notions from control theory like observability also apply for the convolutional codes and could be looked at more closely as I have chosen to look at here for controllability. Essentially, what observability implies is that for any output sequence, if the observability condition checks, then we would be able to compute exactly the corresponding state sequence that produced the output sequence. It is not clear that the old formulations allow such an easy procedure or even any methods to perform similar computations.

\section{Optimal Decoding using the Bowyer Decoder}
Now, in this section I derive the Bowyer Decoder which takes advantage of the reformulation of the convolutional encoder as a discrete-time LTI control system. The key insight is that the dynamics can be coupled with a time-varying cost model. Furthermore, because my decoder combines control systems with communications it admits further extensions to convolutional coding theory. 

Some example applications and future research directions are presented briefly now. The ability to add more cost terms to the model would be easier in my formulation as well as incorporating realistic constraints. Another interesting possibility is to find which output sequence is minimally closest to the optimal state sequence that also requires the least control effort could be found from my reformulation and subsequent decoding algorithm. Along this line of inquiry one could quickly be able to find the $k$th most likely state sequence, etc. 

Now, let us apply DP using the principle of optimality to find the minimizing control input and state as well as the cost-to-go from all states and for all stages. The number of stages $N$ will equal the length of the received sequence $\underbar{r}=\{r_0,r_1,...,r_n\}$ divided by the inverse of the rate $R$. The DP goal will be to find the minimizing output sequence $\underbar{y} = \{y_0,...,y_n\}$ that will come from optimizing the terminal stage costs, moving backwards in time and minimizing all immediate stage costs and cost-to-go values. A final remark: while all encoding operations are done using mod 2 arithmetic, any other vector arithmetic in conjunction with the received vector $r$ is carried out in the field of real numbers. 

First, start with the terminal stage:

\begin{equation*}
    \begin{split}
        J_N(y_N,r_N) &= \min\limits_{(x_N,u_N)} [ || y_N - r_N||^2]\\
            &= \min\limits_{(x_N,u_N)}[||h(x_N,u_N) - r_N||^2]\\
            &= \min\limits_{(x_N,u_N)}[||Cx_N + Du_N - r_N||^2].\\
    \end{split}
\end{equation*}

Hence, our optimal policy will be a function of the control, state and received value at some stage. We can view our "state" as the encoded output.  We can find the optimal policy for terminal stage $N$:

\begin{equation*}
    \begin{split}
    \mu^*_N &= (x^*_N,u^*_N) \\
            &= \arg\min\limits_{(x_N,u_N)} [ || y_N - r_N||^2]\\
            &= \arg\min\limits_{(x_N,u_N)}[||h(x_N,u_N) - r_N||^2]\\
            &= \arg\min\limits_{(x_N,u_N)}[||Cx_N + Du_N - r_N||^2].\\
    \end{split}
\end{equation*}

Next, we complete the new algorithm by working out the arbitrary stage $k\in \{0,...,N-1\}$.

The time-varying stage cost will be the euclidean squared distance between the received data at that stage, and the encoded output for that stage, i.e., at stage $k$,
\begin{equation*}
    \begin{split}
        c_k(y_k,r_k) &= ||y_k-r_k||^2\\
            &= ||h(x_k,u_k) - r_k||^2\\
            &= ||Cx_k + Du_k - r_k||^2.
    \end{split}
\end{equation*}

Now, I will write the form of the DP equation for the general stage $k$. $u_{k+1}$ will be the optimal action for the next stage already determined from the previous stage of DP:
\begin{footnotesize}
\begin{equation*}
    \begin{split}
        J_k(y_k,r_k) &= \min\limits_{(x_k,u_k)}[c_k(y_k,r_k) + J_{k+1}(y_{k+1},r_{k+1})]\\
            &= \min\limits_{(x_k,u_k)}[c_k(h(x_k,u_k),r_k) + J_{k+1}(h(f(x_k,u_k), u_{k+1}),r_{k+1})]\\
            &=\min\limits_{(x_k,u_k)}[c_k(Cx_k+Du_k,r_k) + J_{k+1}(C[Ax_k+bu_k]+ Du^*_{k+1},r_{k+1})],\\
    \end{split}
\end{equation*}
\end{footnotesize}
where $x_{k+1} = f(x_k,u)$ is the updated state using the state space equation parameters $A,B$. Also, $y_{k+1} = h(x_{k+1}, u_{k+1})$ is the updated output making use of the output equation parameters $C,D$. 

After expanding the immediate stage cost,
\begin{small}
$$J_k(y_k,r_k) = \min\limits_{(x_k,u_k)}[||Cx_k+Du_k-r_k||^2 + J_{k+1}(C[Ax_k+bu_k]+ Du^*_{k+1},r_{k+1})]$$
\end{small}

One final simplification we can make is to eliminate $r_k$ from the state space of our cost-to-go functions. It should be viewed as side information we have at any given stage. In fact, the notation $J(y_k|r_k)$ may be more suggestive or intuitive to the underlying decoding. This is more conceptually correct, because our cost-to-go function should be conditioned on whatever the received data is for that stage. Thus, the final Bowyer Decoding equation is: 
\begin{small}
$$ J_k(y_k|r_k) = \min\limits_{(x_k,u_k)}[||Cx_k+Du_k-r_k||^2 + J_{k+1}(C[Ax_k+bu_k]+ Du^*_{k+1}|r_{k+1})].$$
\end{small}

Compressing notation some more:
$$J_k(y_k|r_k) = \min\limits_{(x_k,u_k)}[c_k(x_k,u_k,r_k) + J_{k+1}(h(f(x_k,u_k),u^*_{k+1})|r_{k+1})],$$ and
$$\mu^*_k = \arg\min\limits_{(x_k,u_k)}[c_k(x_k,u_k,r_k) + J_{k+1}(h(f(x_k,u_k),u^*_{k+1})|r_{k+1})].$$

This is the complete decoding procedure for any given convolutional code using the Bowyer Decoder with the convolutional encoder reformulated as an LTI control system. The number of stages is determined by the length of the received vector and the rate of the code. The Bowyer Decoder works on received data, and constructs the optimal state and input sequence. From these optimal states and inputs, the optimal output sequence  is constructed backwards in time using Dynamic programming making full use of the LTI formulation of the encoder. This is how the optimal codeword is obtained using the Bowyer Decoder. Also, I hope this paper can spur more results between the control system communities and communication engineers working in the research areas contained in error control coding.

\bibliographystyle{unsrt}
\bibliography{main}

\begin{thebibliography}{1}

\bibitem{Viterbi67}
Andrew Viterbi.
\newblock Error bounds for convolutional codes and an asymptotically optimum
  decoding algorithm.
\newblock {\em IEEE transactions on Information Theory}, 13(2):260--269, 1967.

\bibitem{Viterbi71}
Andrew Viterbi.
\newblock Convolutional codes and their performance in communication systems.
\newblock {\em IEEE Transactions on Communication Technology}, 19(5):751--772,
  1971.

\bibitem{Viterbi_personal}
Andrew Viterbi.
\newblock A personal history of the viterbi algorithm.
\newblock {\em IEEE Signal Processing Magazine}, 23(4):120--142, 2006.

\bibitem{Forney_personal}
G~David Forney~Jr.
\newblock The viterbi algorithm: A personal history.
\newblock {\em arXiv preprint cs/0504020}, 2005.

\bibitem{Bert17}
Dimitri~P Bertsekas.
\newblock {\em Dynamic programming and optimal control}, volume~1.
\newblock Athena scientific Belmont, MA, 1995.

\end{thebibliography}
\end{document}